\documentclass[a4paper]{article}

\usepackage{INTERSPEECH2022}

\usepackage{multirow}
\usepackage{booktabs}
\usepackage{cite}
\usepackage[urlcolor=blue]{hyperref}
\usepackage{url}
\usepackage{bm}
\usepackage{graphicx}
\usepackage{subfigure}
\title{Towards Improving the Expressiveness of Singing Voice Synthesis with BERT Derived Semantic Information}
\name{Shaohuan Zhou$^{1,\dagger}$\thanks{$^{\dagger}$Work conducted when the first author was intern at Huya Inc.}, Shun Lei$^1$, Weiya You$^1$, Deyi Tuo$^2$,Yuren You$^2$, Zhiyong Wu$^{1,3,*}$\thanks{* Corresponding authors.},  Shiyin Kang$^{4,*}$, Helen Meng$^{1,3}$}
\address{
    $^1$ Shenzhen International Graduate School, Tsinghua University, Shenzhen, China\\
    $^2$ Huya Inc., Guangzhou, China\\
    $^3$ The Chinese University of Hong Kong, Hong Kong SAR, China\\
    $^4$ XVerse Inc., Shenzhen, China
}
\email{\{zhoush21, leis21, ywy19\}$@$mails.tsinghua.edu.cn, 
        tuodeyi$@$huya.com,
        markyouyr$@$gmail.com,
        zywu$@$sz.tsinghua.edu.cn,
        kangshiyin$@$xverse.cn,
        hmmeng$@$se.cuhk.edu.hk
}

\begin{document}

\maketitle
\begin{abstract}
  This paper presents an end-to-end high-quality singing voice synthesis (SVS) system that uses bidirectional encoder representation from Transformers (BERT) derived semantic embeddings to improve the expressiveness of the synthesized singing voice.
  Based on the main architecture of recently proposed VISinger, we put forward several specific designs for expressive singing voice synthesis.
  First, different from the previous SVS models, we use text representation of lyrics extracted from pre-trained BERT as additional input to the model. 
  The representation contains information about semantics of the lyrics, which could help SVS system produce more expressive and natural voice. 
  Second, we further introduce an energy predictor to stabilize the synthesized voice and model the wider range of energy variations that also contribute to the expressiveness of singing voice. 
  Last but not the least, to attenuate the off-key issues, the pitch predictor is re-designed to predict the real to note pitch ratio. 
  Both objective and subjective experimental results 
  indicate that the proposed SVS system can produce singing voice with higher-quality outperforming VISinger\footnote{Samples: \href{https://thuhcsi.github.io/interspeech2022-expressive-svs/}{https://thuhcsi.github.io/interspeech2022-expressive-svs/}} .
\end{abstract}
\section{Introduction}


Singing voice synthesis (SVS) system aims to generate singing voice from the given musical scores that sounds like human singing. 
Unlike text-to-speech (TTS) which only takes text as input, SVS needs to generate voice with correct pronunciation while considering the given pitch and tempo information from the musical score. 
At present, SVS technique is a key component in various applications such as virtual avatars and smartphone assistants and will have more application scenarios in the near future with the development of Metaverse. 
Singing voice and speech share several similar characteristics because they are both produced by the same human voice production system. 
However, because of the distinctive vocal production process, singing voice owns some unique characteristics such as vibration and decrescendo of the long tones which make SVS synthesizing human-like singing voice challenging.

Deep learning methods have achieved tremendous success on various voice generation tasks such as TTS, voice conversion and so on. 
In recent years, a two-stage model consisting of acoustic model and vocoder is the mainstream architecture of SVS system\cite{ren2020deepsinger, zhang2020durian, blaauw2020sequence, hono2018recent}. 
Each of the models at two-stage architecture has been developed independently. Despite the two-stage model having made progress in some tasks, it cannot conceal the problems with its structure, e.g., the mismatch problem due to separate training of neural acoustic model and neural vocoder. To alleviate the two-stage mismatch problems, several works such as FastSpeech2s\cite{ren2020fastspeech}, EATS\cite{JeffDonahue2021EndtoendAT} and VITS\cite{JaehyeonKim2021ConditionalVA} are proposed to be trained in an end-to-end manner recently.
By adopting variational inference augmented with normalizing flows and adversarial training process, VITS\cite{JaehyeonKim2021ConditionalVA} can generate more natural-sounding audio.
Inspired by VITS, the first end-to-end SVS system named VISinger\cite{zhang2021visinger} is proposed recently,
which follows the main architecture of VITS but makes substantial improvements to the prior encoder to produce high-quality singing voices.

%


Although existing SVS systems are capable of producing high quality singing voices, there is still a significant gap between the synthesized songs and those sung by human. 
Existing SVS systems such as ByteSing\cite{gu2021bytesing} and HifiSinger\cite{chen2020hifisinger} can synthesize singing voices that are exactly the same as musical scores, but they still sound somewhat mechanical. 
When people are singing, they will incorporate various details such as emotions and variations in intensity to enhance the infectiousness and expressiveness of the singing voice, based on the semantic information of the lyrics.
To improve the expressiveness of synthesized speech, several efforts have been devoted in the TTS community\cite{hayashi2019pre, xiao2020improving, zhang2021extracting}to incorporate semantic information from pre-trained representations such as BERT\cite{devlin2018bert} as additional input to the TTS model, achieving promising results.


Inspired by these works, in this paper, we build upon VISinger and propose an high quality end-to-end SVS system with the BERT derived semantic embedding and energy predictor to improve the expressiveness of synthesized singing voice.
To the best of our knowledge, this paper is the first work to utilizes the semantic information from lyrics in SVS. 
In particular, we propose an approach that relies on BERT to extract the semantic information from lyrics as an additional input to our model.  
We believe that the semantic embedding of lyrics contains useful information such as the type of word and the meaning of lyrics which will guide the SVS model to synthesize more natural and expressive singing voice.
The expressiveness of the singing voice can be reflected in the abundant singing voice intensity and emotion variation.
Aiming to model the frequently changing energy, we use a singing adaptor that contains an energy predictor to improve the expressiveness of synthesized singing voice.
Furthermore, compared with VISinger which directly predicts the pitch of singing voice, our method predicts the ratio of given note pitch with the aims to synthesize more natural singing voice and avoid the out-of-tune issue. 
Experiments on an open source Mandarin singing corpus named Opencpop show that the proposed method outperforms the VISinger.

\section{Methodology}
\begin{figure}[!tb]
	\centering
	\includegraphics[width=0.9\linewidth, height=1.0\linewidth]{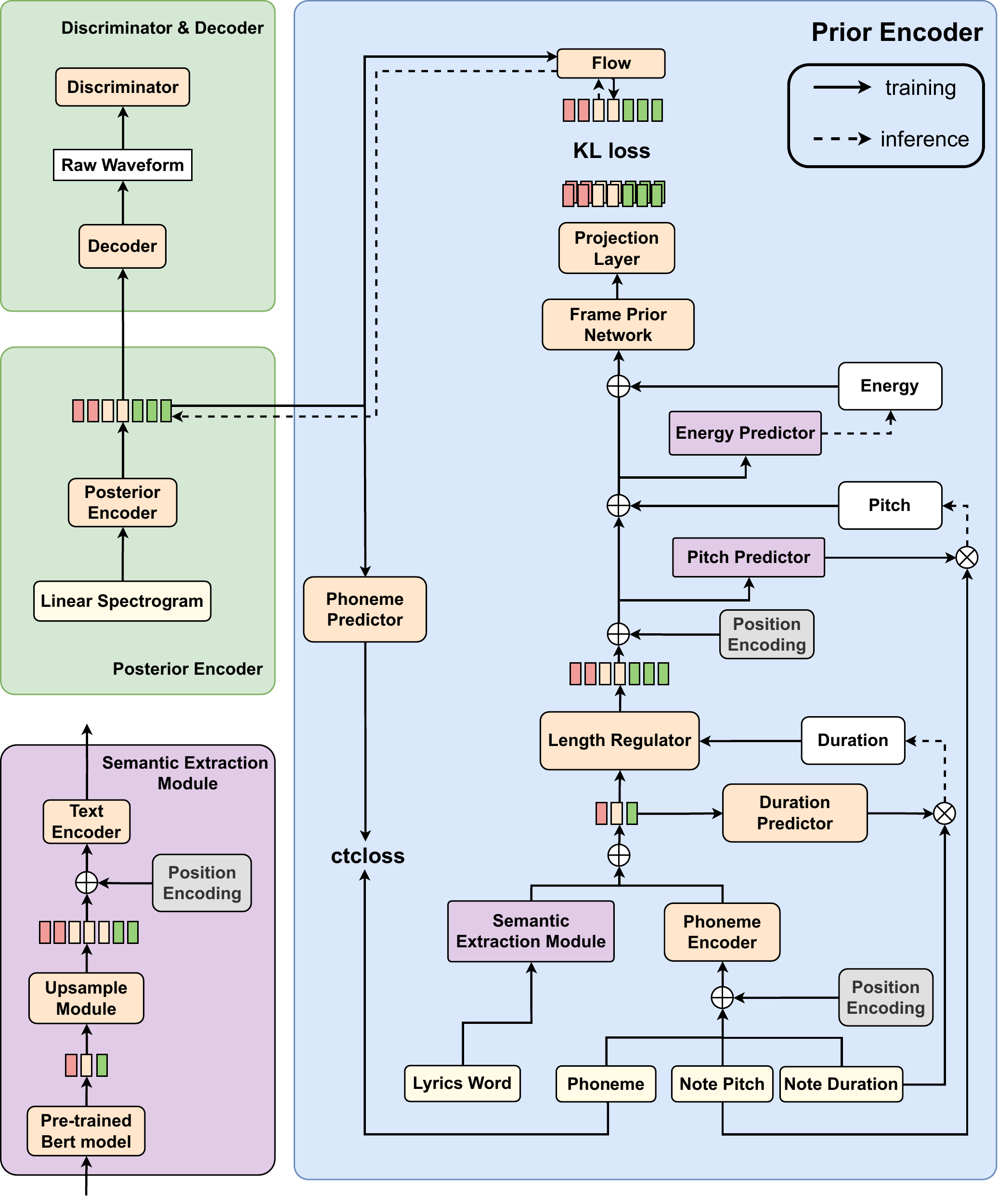}
	\caption{The structure of the proposed SVS system.}
	\label{fig:architecture}
\end{figure}

Following VISinger, as illustrated in Fig.\ref{fig:architecture}, the proposed model consists of several modules: a posterior encoder, a prior encoder and a decoder. 
The posterior encoder takes raw waveform as input, converts it into linear spectrogram and predicts the latent representation
$z$. 
The prior encoder estimates the prior distribution $p(z|c,s)$ of the latent representation $z$ given the music score $c$ and semantic information $s$. 
During inference, the prior encoder takes music score and semantic information as input to predict the frame-level features.
The decoder takes the latent variables generated by the posterior encoder as input during training or the output of the prior encoder during inference. 

In the prior encoder, to utilize the semantic information of lyrics to improve the expressiveness of synthesized singing voice, we introduce a semantic information extraction module to obtain the semantic embedding of lyrics.
Furthermore, we add a singing adaptor that contains a pitch predictor and an energy predictor to stabilize the synthesized voice. 

\subsection{Semantic extraction module}

\subsubsection{Text-context extraction network}
In the prior encoder, we use a pre-trained BERT model to extract the semantic information from the given lyrics. 
The BERT model is a deep bi-directional Transformer\cite{vaswani2017attention} which considers both past and future contexts to extract the text-context as the semantic information.
We hypothesize
that the semantic information contains useful information to guide the SVS model synthesize more natural and expressive singing voice. The BERT model takes the word-level Chinese characters as input and outputs the word-level text embeddings.


\subsubsection{Upsample module}
To combine word-level semantic embedding with phoneme-level output of the phoneme encoder, we adopt upsample module to extend the word-level semantic embedding to phoneme-level by means of replication.
However, the implementation of this upsample module cannot simply be based on the number of Pinyin corresponding to a word, since the phoneme sequence provided by the dataset contains repeated rhymes and some symbols like SP(Silence) and AP(Aspirate) that are not phonemes.
We need to get the number of phonemes corresponding to each word, know the number of times those phonemes have been repeated, and record the index of SP and AP.

Given a text sequence $T = (T_1,T_2,...,T_n)$, we use the Pypinyin module to obtain the corresponding Pinyin of each word.
We then use the pinyin-to-phonemes dictionary provided with the dataset to convert the Pinyin of each word into corresponding phonemes and record the number of phonemes for each Pinyin
as 
$N_1=(P_1,P_2,...,P_n)$.
Then the text sequence $T$ is extended into $T^{'}$according to $N_{1}$.
We further traverse through each of the syncopated utterances to record the number of times each phoneme is repeated in the sequence as $N_{2}=(R_1,R_2,...,R_m)$, where $m$ indicates the length of extended text sequence $T^{'}$.
Then the sequence $T^{'}$ is extended into $T^{''}$ according to $N_{2}$.
After that, we record the index of SP and AP and insert the all-zero vector in the corresponding position of the sequence.
After the above steps, the word-level semantic embeddings are extended to phoneme-level.

\subsubsection{Text encoder}
Since the word-level semantic embedding is augmented with a copy operation in the upsample module, we use a text encoder to model the phoneme-level semantic embedding. As illustrated in Fig.\ref{fig:architecture}, we add position encoding to the upsampled semantic embedding and send it into the text encoder.  
The text encoder is a Transformer encoder with 6 FFT blocks and an additional linear layer is added to text encoder 
to project the hidden representations so as to make whose channel numbers compatible with the output of phoneme encoder.


\subsection{Pitch predictor}
Since people sing with a natural and expressive voice that fluctuates up and down around the note pitch, we introduce a pitch predictor to predict the pitch deviation between the synthesized voice and the note pitch. We follow the architecture of F0-predictor in VISinger. 

Different from the F0-predictor which directly predicts the pitch of the synthesized voice, our pitch predictor predicts the ratio $r$ of synthesized pitch to the note pitch, then the ratio $r$ is multiplied with the note pitch to obtain the synthesized pitch. 
In this way, the pitch predictor only needs to predict the deviation of note pitch instead of final pitch, which makes pitch predictor more robust to unseen data.
Specifically, given note pitch $p$ and ratio $r$, the predicted pitch is defined as follows: 
\begin{align}
    f &= 440 \times 2 ^ {(p-69)/12} \\
    LF0 &= logf \\
    \hat{LF0} &= r \times LF0
\end{align}
where $f$ is the corresponding frequency of pitch id $p$. Then the predicted pitch is quantified and converted into pitch embedding by an embedding layer. Finally, the pitch loss can be expressed as follows:
\begin{align}
    L_{pitch} &= \left \| L_{wav} - \hat{LF0} \right\|_2
\end{align}
where $L_{wav}$ represents the logarithmic pitch of the given waveform.

\subsection{Energy predictor}
Compared with speech, singing voice have a wider range of energy variations.
It is vital to model the energy changes of the singing voice to improve the expressiveness of synthesized singing voice. 
We further introduce an energy predictor to predict the energy of singing voice.

Specifically, the energy predictor consists of a 2-layer 1D-convolutional network with ReLU activation, each followed by a layer normalization and a dropout layer.
To project the hidden states into the output sequence, an extra linear layer is added in the energy predictor.
Following \cite{ren2020fastspeech}, we compute the L2-norm of the amplitude of each short-time Fourier transform (STFT) frame as the ground-truth energy. 
The energy predictor takes the frame-level hidden sequence as input and predicts the energy of each frame. The predicted energy is converted into logarithmic domain for ease of prediction.
Given the ground-truth energy $E$ and the predicted energy $\hat{E}$, the energy loss is expressed as:
\begin{gather}
    L_{energy} = \left \| E - \hat{E} \right\|_2
\end{gather}
Besides, we quantize the energy of each frame to 256 possible values uniformly, encode it into energy embedding and add it with the pitch embedding and the expanded hidden sequence.

\section{Experiments}
\subsection{Datasets}
In this paper, we use a recently public Mandarin singing corpus named Opencpop\cite{wang2022opencpop} for SVS experiments. 
The Opencpop corpus consists of 100 popular Mandarin songs performed by a female professional singer. The audio of Opencpop are all recorded at a sampling rate of 44,100Hz. We downsample them into 24,000Hz with 16-bit quantization. All singing recordings have been phonetically annotated with utterance, note and phoneme boundaries. The final dataset is cut into 3,756 utterances with a total of about 5.2 hours. Among them, 3,550 utterances are randomly selected for training while the rest 206 utterances for validation and test.

\subsection{System configuration}
The following three systems are constructed for evaluating the performance of our proposed systems. 

\textbf{VISinger:} 
A complete end-to-end singing voice synthesis system that adopts VAE-based posterior encoder augmented with normalizing flow-based prior encoder and adversarial decoder. We implement VISinger based on VITS\cite{JaehyeonKim2021ConditionalVA}. In VISinger, the embedding dimension of the input features is 192. Each note pitch is converted into a pitch ID following the MIDI standard\cite{midi}, while the note duration is converted into frame count by the sampling rate, window length and hop length.
Specifically, given note duration $dur$, sampling rate $sr$, window length $wl$ and hop length $hl$, the frame count $frame$ is calculated as:
\begin{align}
    frame &= [(dur * sr) - wl]/hl + 1
\end{align}
All the embeddings of the input features are summed and fed into the model. The text encoder contains 6 FFT blocks, and the duration predictor consists of 3 one-dimensional convolutional networks. The note normalization is also adopted in VISinger to predict the phoneme duration based on note duration. 

\textbf{Proposed:} The modified VISinger architecture that adopts all the contributions introduced in the paper. The dimension of text embedding derived from BERT is 768 and converted into 192 by the additional linear layer in text encoder. Besides, the energy predictor predicts the energy of each frame and converts it into energy embedding with 192 dimensions that are the same as the pitch predictor.
The rest of the hyperparameter settings are consistent with those in VISinger.

\textbf{Proposed with reverse SEM:} The proposed system that use a reverse semantic extraction module (SEM). To justify the structure of the proposed semantic extraction module, we replace 
it
with a reversed semantic extraction module (SEM). Specifically, the reversed SEM reverses the order of upsample module and text encoder. The word-level 
semantic representation
is firstly encoded by the text encoder and then extended into phoneme-level features by the upsample module.




The proposed system and VISinger are trained up to 200k steps on 8 Nvidia Tesla A100, and the batch size for each GPU is 4. We use the AdamW\cite{loshchilov2017decoupled} optimizer with $\beta_1$ = 0.8, $\beta_2$ = 0.99, $\epsilon$ = $10^{-9}$. The initial learning rate is set to $1e-4$ and with a learning decay of $0.999875$.

\subsection{Experimental results}

We first conduct the mean opinion score (MOS) test to evaluate the naturalness and expressiveness of the synthesized singing voice by subjective listening tests.
For each system, we prepare 10 audio samples, with each sample within about 10 seconds. 
25 Chinese native speakers are recruited as subjects to rate the given audios on a scale from 1 to 5 with 1 point interval. 
As shown in Table.\ref{tab:mos}, our method achieves a better MOS of 3.808, exceeding VISinger by 0.492, which demonstrates the effectiveness of our method over VISinger.




\begin{table}[th]
\footnotesize
\renewcommand{\arraystretch}{1.0}
  \caption{Experimental results in terms of subjective mean opinion score (MOS) with 95\% confidence intervals and three objective metrics}
  \setlength{\tabcolsep}{1mm}{
  \label{tab:mos}
  \centering
  \begin{tabular}{l|c|c|c|c} 
    \toprule
    \textbf{Model} &\textbf{MOS}&\textbf{F0 MAE} & \textbf{Dur MAE}& \textbf{Energy MAE} \\
    \midrule
    VISinger & $3.316\pm0.101$& $13.193$ &$8.451$&$21.974$  \\
    Proposed & \bm{$3.808\pm0.085$}&  \bm{$12.897$}&\bm{$6.682$}&\bm{$16.523$}  \\
    \midrule
    Recording & $4.876\pm0.043$ & - & - & -\\
    \bottomrule
  \end{tabular}}
\end{table}

\begin{figure*}[htbp]
	\centering
	\subfigure[Ground truth] {\includegraphics[width=.3\textwidth]{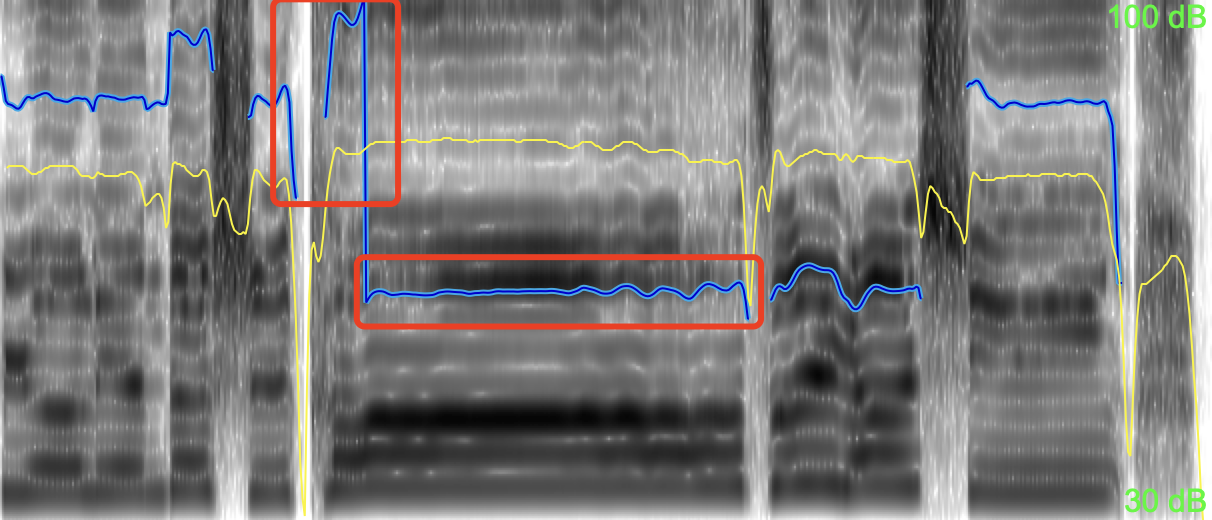}}
	\subfigure[Proposed] {\includegraphics[width=.3\textwidth]{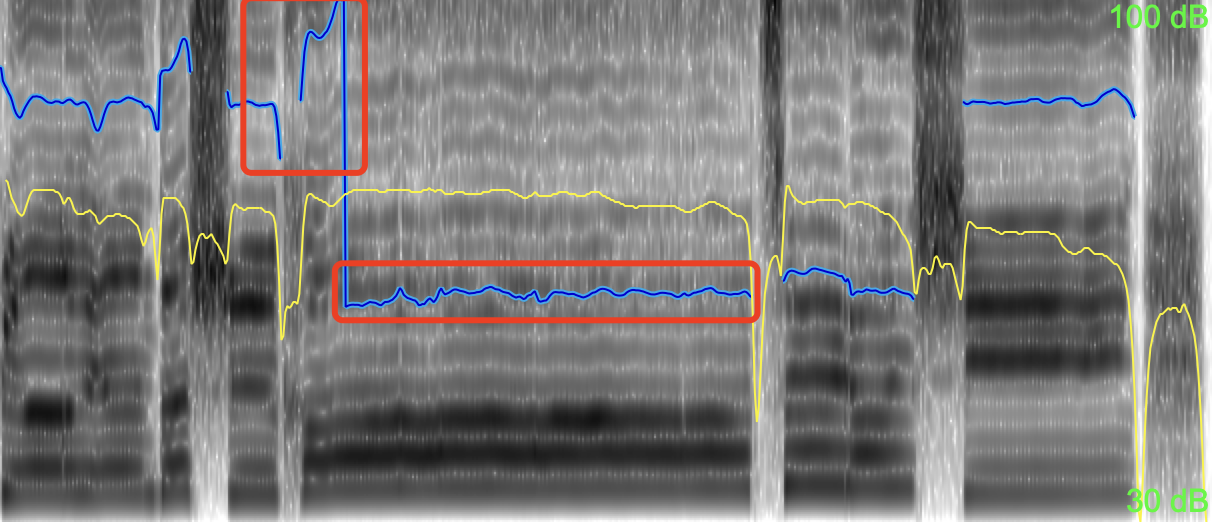}}
	\subfigure[VISinger] {\includegraphics[width=.3\textwidth]{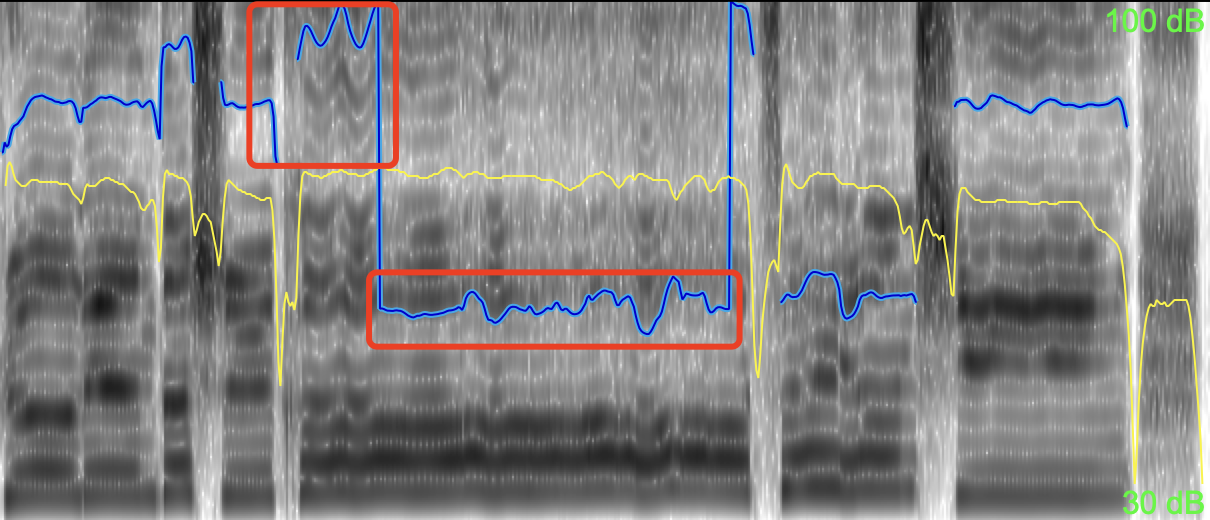}}
	\caption{Visualizations of spectrograms, energy and pitch contours in three systems: Ground truth, Proposed, and VISinger. The yellow line represents the energy contour, while the blue line represents the pitch contour.}
	\label{case}
\end{figure*}




We further calculate three objective metrics including F0 mean absolute error (F0 MAE), duration mean absolute error (Dur MAE), and energy mean absolute error (Energy MAE).
The objective results in Table.\ref{tab:mos} indicate that the synthesized singing voice from the proposed method can achieve more precise pitch prediction, which indicates the effectiveness of the re-designed pitch predictor.
Moreover, the proposed method achieves better results in Dur MAE and Energy MAE compared to VISinger, showing our method has a stronger ability to synthesize 
singing voice closer to the ground truth.



\begin{figure}[!htb]
	\centering
	\includegraphics[width=1\linewidth, height=0.4\linewidth]{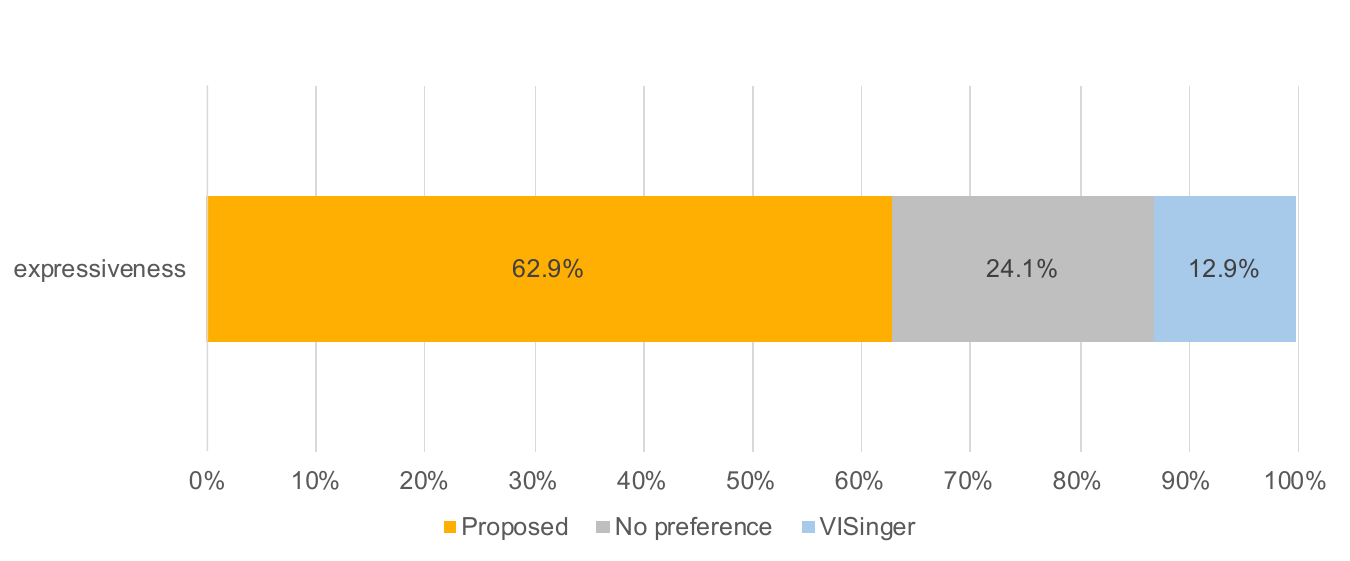}
	\caption{Result of the ABX preference test for expressiveness.}
	\label{fig:abx}
\end{figure}


ABX preference test is also carried out to ask subjects to give their preferences in terms of expressiveness between a pair of singing voices synthesized by the proposed method and VISinger.
We randomly select 10 segments from the test set and recruit 14 native Chinese speakers to vote for the given audios. The result is shown in Fig.\ref{fig:abx}.
Compared to VISinger, which does not use semantic information, the singing voice generated by the proposed method is supported by 62.9\% of the subjects and exceeds VISinger by $50\%$.
It demonstrates that our proposed semantic extraction module can help the SVS model generate a more expressive singing voice.

\subsection{Ablation study}

To demonstrate the effectiveness of different contributions in our proposed method, including the energy predictor and the semantic extraction module, we carry out three ablation studies. 
We employ comparison mean opinion score (CMOS) to compare the synthesized singing voices in terms of naturalness and expressiveness.
As shown in Table.\ref{tab:cmos}, the removal of energy predictor result in -0.868 CMOS, indicating that the energy predictor improves the stability of the synthesized singing voice and thus the naturalness of the synthesized voice. 
Removing of the semantic extraction module results in -0.456 CMOS, showing that the proposed module can effectively extract the semantic information from the given lyrics and the SVS model can utilize it to synthesize more expressive singing voice.

Furthermore, to demonstrate the rationality of our semantic extraction module, we redesign the semantic extraction module
as reversed SEM. 
As shown in the last line in Table.\ref{tab:cmos}, the use of reversed SEM structure results in -0.108 CMOS, demonstrating the 
validity and effectiveness
of the proposed semantic extraction module.

\begin{table}[th]
\renewcommand{\arraystretch}{1.0}
  \caption{Ablation study results about the effect of each component on the naturalness and expressiveness of singing voice.}
  \label{tab:cmos}
  \centering
  \begin{tabular}{l|c} 
    \toprule
    \textbf{Model} &\textbf{CMOS} \\
    \midrule
    Proposed & $0.000$ ~~~\\
     - energy predictor & $-0.868$ ~~~  \\
     - semantic extraction module & $-0.456$ ~~~  \\
    Proposed with reversed SEM & $-0.108$ ~~~  \\
    \bottomrule
  \end{tabular}
\end{table}

\subsection{Case study}
To explore the impact of the aforementioned contributions, we further conduct a case study to synthesize a singing voice segment from the test set with sudden pitch changes and long tones. 
We synthesize it by our proposed method and VISinger, and the ground-truth recording is also provided for comparison. 
As shown in Fig.\ref{case}, pitch and energy curves are marked with blue and yellow lines respectively in our plotted spectrogram. 
In the long tones circled in the middle red box, the singing voice synthesized by the proposed method has a flat pitch, while the singing voice synthesized by VISinger has irregular fluctuations, making it sound incoherent.
This can be attributed to our re-designed pitch predictor, which makes our model predict pitch more accurately.
In the red box circled in the upper-left corner, the pitch of our proposed method changes more similar to that of the ground-truth, with a more pronounced pitch dip before the sudden pitch change, followed by a sliding rise to the highest pitch.
It can be seen that our proposed method can synthesize singing voice with a more natural pitch compared to VISinger.
Furthermore, with the help of the energy predictor, the synthesized singing voice of the proposed method and ground-truth recording both have a slight fade at the end of the long tones, while that of VISinger does not.
The result of the case study demonstrates the improvement of our proposed method in synthesizing expressive singing voice.
\section{Conclusions}

In this paper, we present a high-quality singing voice synthesis system. 
The proposed system is built upon VISinger - the recently proposed end-to-end singing voice synthesis system. 
To improve the expressiveness of synthesized singing voice, we introduce several contributions, including 
a semantic extraction module, 
an energy predictor 
and a re-designed pitch predictor. 
Experiments on a recent public Mandarin singing corpus named Opencpop show that the proposed system outperforms the VISinger in both subjective evaluation and objective indicators.

In the future, we will continue to explore ways to improve the expressiveness of our synthesized singing voice to further bridge the gap between synthesized voice and recorded singing voice performed by humans.




\section{Acknowledgements}


This work is supported by National Natural Science Foundation of China (NSFC) (62076144), National Social Science Foundation of China (NSSF) (13\&ZD189) and Shenzhen Key Laboratory of next generation interactive media innovative technology (ZDSYS20210623092001004).

\newpage

\bibliographystyle{IEEEtran}
\bibliography{references}

\begin{thebibliography}{10}
\providecommand{\url}[1]{#1}
\csname url@samestyle\endcsname
\providecommand{\newblock}{\relax}
\providecommand{\bibinfo}[2]{#2}
\providecommand{\BIBentrySTDinterwordspacing}{\spaceskip=0pt\relax}
\providecommand{\BIBentryALTinterwordstretchfactor}{4}
\providecommand{\BIBentryALTinterwordspacing}{\spaceskip=\fontdimen2\font plus
\BIBentryALTinterwordstretchfactor\fontdimen3\font minus
  \fontdimen4\font\relax}
\providecommand{\BIBforeignlanguage}[2]{{%
\expandafter\ifx\csname l@#1\endcsname\relax
\typeout{** WARNING: IEEEtran.bst: No hyphenation pattern has been}%
\typeout{** loaded for the language `#1'. Using the pattern for}%
\typeout{** the default language instead.}%
\else
\language=\csname l@#1\endcsname
\fi
#2}}
\providecommand{\BIBdecl}{\relax}
\BIBdecl

\bibitem{ren2020deepsinger}
Y.~Ren, X.~Tan, T.~Qin, J.~Luan, Z.~Zhao, and T.-Y. Liu, ``Deepsinger: Singing
  voice synthesis with data mined from the web,'' in \emph{Proceedings of the
  26th ACM SIGKDD International Conference on Knowledge Discovery \& Data
  Mining}, 2020, pp. 1979--1989.

\bibitem{zhang2020durian}
L.~Zhang, C.~Yu, H.~Lu, C.~Weng, C.~Zhang, Y.~Wu, X.~Xie, Z.~Li, and D.~Yu,
  ``Durian-sc: Duration informed attention network based singing voice
  conversion system,'' \emph{arXiv preprint arXiv:2008.03009}, 2020.

\bibitem{blaauw2020sequence}
M.~Blaauw and J.~Bonada, ``Sequence-to-sequence singing synthesis using the
  feed-forward transformer,'' in \emph{ICASSP 2020-2020 IEEE International
  Conference on Acoustics, Speech and Signal Processing (ICASSP)}.\hskip 1em
  plus 0.5em minus 0.4em\relax IEEE, 2020, pp. 7229--7233.

\bibitem{hono2018recent}
Y.~Hono, S.~Murata, K.~Nakamura, K.~Hashimoto, K.~Oura, Y.~Nankaku, and
  K.~Tokuda, ``Recent development of the dnn-based singing voice synthesis
  system—sinsy,'' in \emph{2018 Asia-Pacific Signal and Information
  Processing Association Annual Summit and Conference (APSIPA ASC)}.\hskip 1em
  plus 0.5em minus 0.4em\relax IEEE, 2018, pp. 1003--1009.

\bibitem{ren2020fastspeech}
Y.~Ren, C.~Hu, X.~Tan, T.~Qin, S.~Zhao, Z.~Zhao, and T.-Y. Liu, ``Fastspeech 2:
  Fast and high-quality end-to-end text to speech,'' \emph{arXiv preprint
  arXiv:2006.04558}, 2020.

\bibitem{JeffDonahue2021EndtoendAT}
J.~Donahue, S.~Dieleman, M.~Bińkowski, E.~Elsen, and K.~Simonyan, ``End-to-end
  adversarial text-to-speech,'' in \emph{International Conference on Learning
  Representations}, 2021.

\bibitem{JaehyeonKim2021ConditionalVA}
J.~Kim, J.~Kong, and J.~Son, ``Conditional variational autoencoder with
  adversarial learning for end-to-end text-to-speech,'' \emph{arXiv: Sound},
  2021.

\bibitem{zhang2021visinger}
Y.~Zhang, J.~Cong, H.~Xue, L.~Xie, P.~Zhu, and M.~Bi, ``Visinger: Variational
  inference with adversarial learning for end-to-end singing voice synthesis,''
  \emph{arXiv preprint arXiv:2110.08813}, 2021.

\bibitem{gu2021bytesing}
Y.~Gu, X.~Yin, Y.~Rao, Y.~Wan, B.~Tang, Y.~Zhang, J.~Chen, Y.~Wang, and Z.~Ma,
  ``Bytesing: A chinese singing voice synthesis system using duration allocated
  encoder-decoder acoustic models and wavernn vocoders,'' in \emph{2021 12th
  International Symposium on Chinese Spoken Language Processing
  (ISCSLP)}.\hskip 1em plus 0.5em minus 0.4em\relax IEEE, 2021, pp. 1--5.

\bibitem{chen2020hifisinger}
J.~Chen, X.~Tan, J.~Luan, T.~Qin, and T.-Y. Liu, ``Hifisinger: Towards
  high-fidelity neural singing voice synthesis,'' \emph{arXiv preprint
  arXiv:2009.01776}, 2020.

\bibitem{hayashi2019pre}
T.~Hayashi, S.~Watanabe, T.~Toda, K.~Takeda, S.~Toshniwal, and K.~Livescu,
  ``Pre-trained text embeddings for enhanced text-to-speech synthesis.'' in
  \emph{INTERSPEECH}, 2019, pp. 4430--4434.

\bibitem{xiao2020improving}
Y.~Xiao, L.~He, H.~Ming, and F.~K. Soong, ``Improving prosody with linguistic
  and bert derived features in multi-speaker based mandarin chinese neural
  tts,'' in \emph{ICASSP 2020-2020 IEEE International Conference on Acoustics,
  Speech and Signal Processing (ICASSP)}.\hskip 1em plus 0.5em minus
  0.4em\relax IEEE, 2020, pp. 6704--6708.

\bibitem{zhang2021extracting}
Y.-J. Zhang and Z.-H. Ling, ``Extracting and predicting word-level style
  variations for speech synthesis,'' \emph{IEEE/ACM Transactions on Audio,
  Speech, and Language Processing}, vol.~29, pp. 1582--1593, 2021.

\bibitem{devlin2018bert}
J.~Devlin, M.-W. Chang, K.~Lee, and K.~Toutanova, ``Bert: Pre-training of deep
  bidirectional transformers for language understanding,'' \emph{arXiv preprint
  arXiv:1810.04805}, 2018.

\bibitem{vaswani2017attention}
A.~Vaswani, N.~Shazeer, N.~Parmar, J.~Uszkoreit, L.~Jones, A.~N. Gomez,
  {\L}.~Kaiser, and I.~Polosukhin, ``Attention is all you need,''
  \emph{Advances in neural information processing systems}, vol.~30, 2017.

\bibitem{wang2022opencpop}
Y.~Wang, X.~Wang, P.~Zhu, J.~Wu, H.~Li, H.~Xue, Y.~Zhang, L.~Xie, and M.~Bi,
  ``Opencpop: A high-quality open source chinese popular song corpus for
  singing voice synthesis,'' \emph{arXiv preprint arXiv:2201.07429}, 2022.

\bibitem{midi}
M.~M. Association, ``Midi manufacturers association,''
  \url{https://www.midi.org}.

\bibitem{loshchilov2017decoupled}
I.~Loshchilov and F.~Hutter, ``Decoupled weight decay regularization,''
  \emph{arXiv preprint arXiv:1711.05101}, 2017.

\end{thebibliography}

\end{document}